\documentclass{article}%
\usepackage{graphicx}
\usepackage{amsmath}%
\usepackage{amsfonts}%
\usepackage{amssymb}

\begin{document}

\title{Brussels-Austin Nonequilibrium Statistical Mechanics in the Later Years: Large
Poincar\'{e} Systems and Rigged Hilbert Space}
\author{Robert C. Bishop$^{a,b}$\\$^{a}$Abteilung f\"{u}r Theorie und Datenanalyse, Institut\\f\"{u}r Grenzgebiete der Psychologie, Wilhelmstrasse\\3a, D-79098 Freiburg, Germany\\$^{b}$Permanent Address: Department of Philosophy,\\Logic and Scientific Method, The London School\\of Economics, Houghton St., London,\\WC2A 2AE, United Kingdom\\Draft}
\date{}
\maketitle

\begin{abstract}
This second part of a two-part essay discusses recent developments in the
Brussels-Austin Group after the mid 1980s. The fundamental concerns are the
same as in their similarity transformation approach (see Part I), but the
contemporary approach utilizes rigged Hilbert space (whereas the older
approach used Hilbert space). While the emphasis on nonequilibrium statistical
mechanics remains the same, the use of similarity transformations shifts to
the background. In its place arose an interest in the physical features of
large Poincar\'{e} systems, nonlinear dynamics and the mathematical tools
necessary to analyze them.


Word count (including notes): 9,547

\end{abstract}

\section{Introduction}

Part I of this essay discussed the earlier similarity transformation approach
to nonequilibrium statistical mechanics of Ilya Prigogine and his coworkers.
This approach, along with that of subdynamics, is perhaps somewhat familiar as
it has received some attention in philosophical literature and was the subject
of Prigogine's well-known book, \textit{From Being to Becoming: Time \&
Complexity in the Physical Sciences} (1980). Part II of this essay focuses on
their more recent and less familiar work on nonequilibrium statistical
mechanics in rigged Hilbert spaces.

It has been argued that no current approaches to microscopic dynamics can
explain or derive the second law of thermodynamics, since it is both necessary
and sufficient for the derivation of the second law from microscopic dynamics
that the dynamics be exact (e.g. Mackey 1992, pp. 98-100; 2002).\footnote{A
dynamics on a state space $\Omega$ with a transfer operator $P_{t}$ is
\textit{exact} if and only if lim$_{t\rightarrow\infty}$ $|P_{t}\rho-\rho
_{eq}|_{L^{1}}=0$ for every initial density $\rho$, where $\rho_{eq}$ is the
unique stationary density (i.e. equilibrium density), $P_{t}$ governs the
dynamics (e.g. Liouville or the Frobenius-Perron operators), and the norm is
in the sense of Lebesgue integrable functions. Among other properties, exact
dynamics are noninvertible and always yield a unique stationary density.}
Although it can be shown that the coarse-grained projection operator arising
from the earlier Brussels-Austin approach yields an exact dynamics, whether
their similarity transformation yields exact dynamics is unknown (Antoniou and
Gustafson 1993; Antoniou, Gustafson and Suchanecki 1998, p. 119).
Nevertheless, one of the crucial claims of the earlier approach was that
trajectory descriptions at the microscopic level and probabilistic
descriptions at the macroscopic level of thermodynamic behavior are related
via a transformation (Part I).

This way of viewing the relationship be trajectory and probabilistic
descriptions is de-emphasized in their more recent work. So the core point is
no longer to derive irreversible thermodynamic behavior from reversible
microscopic descriptions, so much as to argue for the priority of irreversible
macroscopic descriptions for a particular class of systems known as large
Poincar\'{e} systems. However, the core intuitions of the new approach remain
continuous with their earlier work; namely, that irreversibility is
fundamentally dynamical in character and that distributions are ontologically
fundamental explanatory elements for complex statistical systems.

The Brussels-Austin Group's recent work develops a method for constructing a
complete set of eigenvectors for the model equations describing the
thermodynamic approach to equilibrium for Large Poincar\'{e} systems as well
as nonlinear dynamics more generally. This approach reformulates the question
of how to relate reversible trajectory and irreversible probabilistic
descriptions as follows: How can the trajectory dynamics of a large
Poincar\'{e} system (LPS) yield necessary conditions for the thermodynamics
approach to equilibrium and what further mechanisms account for the sufficient
conditions for such behavior?

Large Poincar\'{e} systems are defined and illustrated in \S2 using
nonintegrable Hamiltonians and classical perturbation theory as a way of
motivating some of the key physical and mathematical problems for such
systems. The rigged Hilbert space approach to these systems is outlined in
\S3, and the corresponding time-ordering rule and semigroup operators
governing the dynamics are introduced. Particular details of the approach are
discussed in \S4, where an alternative interpretation of Prigogine's treatment
of trajectories and their relationship to the dynamics of distributions is
developed. Some remarks on probabilistic vs. deterministic dynamics closes the
essay (\S5).

\section{Large Poincar\'{e} Systems and Integrability}

Toward the end of the 19th century, Poincar\'{e} was investigating planetary
motion, among other things. Solving the equations of motion for the solar
system is extremely difficult because all the planets interact with each other
through gravitational forces. One of the questions Poincar\'{e} pursued was
whether there was a suitable way to transform these equations of motion into a
system of equations where the gravitational interaction would vanish and one
could solve the evolution equations for the angle variables of each planet
independently of the others. What Poincar\'{e} showed was that in general such
a transformation was impossible for systems of $N$ mutually interacting
bodies. If a canonical transformation for a system of equations describing a
set of interacting particles that carries the equations into a set where the
interactions vanish exists, then the system is classified as
\textit{integrable}. This means that the original system of equations can be
transformed into one where each particle's angle variable is fully described
by an equation that is independent of any other particle's angle variable.

Poincar\'{e} showed that systems of equations were nonintegrable when they
contained \textit{resonances} between various degrees of freedom. In essence a
resonance is a transient metastable state establishing a narrow, precise
frequency gateway through which energy can be efficiently transferred from one
element of a physical system to another. Physical examples of resonances
include transient bound states produced in particle collisions and transient
intermediates in chemical reactions.

\subsection{Integrable Systems and Classical Perturbation Theory}

In order to make these notions of resonances and nonintegrability more
precise, consider Hamiltonian systems in classical mechanics. While models
with completely integrable Hamiltonians are rare, they are still very useful
in the study of physical systems. For many systems can be modeled using
Hamiltonians of the form%
\begin{equation}
H=H_{0}(\vec{J})+\lambda V(\vec{J},\vec{\alpha})\text{,}%
\end{equation}
where $H_{0}$ is assumed to be completely integrable, $\vec{J}$ represents the
action variables (e.g. generalized momentum vectors), $\vec{\alpha}$ the angle
variables (e.g. generalized coordinate vectors) and $\lambda$ (assumed $\ll1$)
is the coupling coefficient roughly describing the strength of the
interactions through the potential $V$. The question of whether or not a
Hamiltonian system is integrable is equivalent to being able to find a
canonical transformation from the old state space coordinates $(\vec{J}%
,\vec{\alpha})$ to the new coordinates $(\vec{I},\vec{\beta})$ corresponding
to a transformation operator of the form%
\begin{equation}
e^{iF(\vec{I},\vec{\beta})}%
\end{equation}
decoupling all the equations for the angle variables (in essence turning off
all the interactions by making $\lambda$ zero). When such a transformation can
be found, the Hamiltonian is said to be completely integrable and I will refer
to this type of integrability as complete integrability (to be distinguished
from the Brussels-Austin sense of integrability below).

In general one then must proceed using a perturbation method where the
strategy is to find approximate solutions of (1) in terms of $H_{o}(\vec{I})$
plus small perturbations due to $V(\vec{I},\vec{\beta})$. In the course of
standard perturbation analysis of such a model (e.g. Tabor 1989, 89-108),
terms of the form%
\begin{equation}
\frac{V_{n_{i},n_{j},n_{k}...}}{n_{i}\beta_{i}+n_{j}\beta_{j}+n_{k}\beta
_{k}+...}%
\end{equation}
emerge where $i$, $j$, and $k$ are integers labeling the particles,
$V_{n_{i}n_{j}n_{k}}$ represents the Fourier transformed potential, the
$n_{l}$ indicate the (discrete) degrees of freedom of the particles in the
Fourier expansion, and the $\beta_{l}$ can be negative and are often
interpreted as generalized frequencies. Clearly terms like (3) increase
without bounds when the denominator approaches zero. The denominator being
zero represents a resonance. It is the presence of a sufficient number of
these resonances that prevents us from using the standard canonical
transformation techniques to turn the model into a completely integrable
system of equations. For an N-body problem, the resonance condition takes the
form that the finite sum $n_{i}\beta_{i}+n_{j}\beta_{j}+n_{k}\beta
_{k}+...+n_{N}\beta_{N}=0$. In general there are several combinations of
$n_{l}$'s and $\beta_{l}$'s satisfying this condition.

\subsection{Large Poincar\'{e} Systems}

First consider an integrable Hamiltonian for a system with two degrees of
freedom. The state space trajectories will then be confined to the surfaces of
nested tori, where each surface corresponds to a different combination of the
values of the two constants of the motion. Now add perturbations $\lambda V$
to this Hamiltonian where $\lambda\ll1$. If the perturbations leave the
Hamiltonian integrable, then the model dynamics are not appreciably affected.
In contrast, if the perturbations render the Hamiltonian nonintegrable (e.g.
resonance phenomena), then these periodic orbits will be disrupted because
such perturbations are as physically important as the unperturbed orbits of
the integrable part of the model, due to the transfer of energy involved. The
KAM theorem specifies the conditions under which tori associated with
quasi-periodic trajectories survive and constitute the majority of motions
realized in state space, so that most regions in state space for nonintegrable
models close to integrable models show stable nonperiodic orbits (e.g. Hilborn
1994, 337-9).

There are two types of fixed points for the state space trajectories in
Hamiltonians of the form (1): elliptic and hyperbolic (saddle points).
Elliptic fixed points correspond to stable periodic orbits which are disrupted
by resonances. Hyperbolic fixed points present complex behavior: trajectories
exhibiting sensitive dependence on initial conditions and which wander
erratically over large regions of state space. These structures also exhibit
self-similarity. The chaotic behavior in Hamiltonian systems is similar to
chaotic behavior in dissipative systems. However, since Hamiltonian systems do
not contract to some fixed point as do dissipative systems, orbits near
hyperbolic fixed points will become unstable leading to exponentially
diverging trajectories. It should be pointed out that stable and chaotic
orbits can coexist simultaneously in state space.

Large Poincar\'{e} systems are of interest to Prigogine and coworkers.
Consider a typical SM Hamiltonian of the form%
\begin{equation}
H(p,q)=\sum_{i=1}^{N}\frac{\vec{P}_{i}^{2}}{2m_{i}}+\lambda\sum_{j>i}%
^{N}V(|\vec{q}_{i}-\vec{q}_{j}|)\text{,}%
\end{equation}
where $\vec{q}$ and $\vec{p}$ are $N$-component vectors representing
generalized coordinates and momenta respectively, and the system is in a large
box with volume $L^{3}$. The Brussels-Austin group is interested in ``large''
systems, meaning they work in the limit $L^{3}\rightarrow\infty$ (the number
of particles $N$ may be finite or infinite). A LPS is obtained when the system
is large and the number of degrees of freedom of the system tends to infinity.
An example of a LPS with a finite number of particles would be a finite number
of charges interacting with an electromagnetic field, while an example with an
infinite number of particles would be the \textit{thermodynamic limit}
($L^{3}\rightarrow\infty$, $N\rightarrow\infty$, $N/L^{3}$ finite). Such
systems possess ``continuous sets of resonances''. By continuous sets of
resonances, the Brussels-Austin Group means that in the Fourier transformed
representation, the eigenfrequencies are continuous functions of the wave
vector $k$, so that the summation operations over terms like (3) must be
replaced by integrals and the denominators of such terms can be arbitrarily
close to zero.

The resonance condition for a continuous set of resonances for a LPS in the
context of perturbation theory takes the form%
\begin{equation}
\int\int b\beta dbd\beta=0\text{,}%
\end{equation}
where $b$ (representing degrees of freedom) and $\beta$ are continuous
functions defined over the real numbers. Under condition (5) motion will not
even be quasi-periodic so that variables have a continuous
spectrum.\footnote{As Koopman and von Neumann first pointed out, for dynamical
systems with continuous spectra, `the states of motion corresponding to any
set become more and more spread out into an amorphous everywhere dense
\textit{chaos}. Periodic orbits, and such like, appear only as very special
possibilities of negligible probability' (Koopman and von Neumann 1932, p.
261). This is generally acknowledged to be the first reference to the term
``chaos'' in the context of dynamics.} No canonical transformation exists that
can turn these LPS models into completely integrable models (Prigogine et al.
1991, pp. 6-7). Such models exhibit the type of randomness associated with
mixing, K-flows and Bernoulli systems, but are usually interpreted as
deterministic.\footnote{The Baker's transformation is a favorite model of a
deterministic random system for the Brussels-Austin Group (Part I). The
equations are reversible, deterministic and conservative, yet the mapping
turns out to have the Bernoulli property (randomness of a coin toss).}

As an example of a LPS, imagine a gas containing an infinite number of
particles continually undergoing collisions, where the collision processes
never cease. A more realistic example is an electromagnetic oscillator with
frequency $\omega_{osc}$ interacting with an electromagnetic field. The field
has an infinite number of degrees of freedom and the frequency $\omega_{k}$ of
the field varies continuously with \textit{k}, giving rise to an infinite
number of resonances. Continuous resonances like those in LPS are involved in
fundamental phenomena such as absorption and emission of light, decay of
unstable particles and the scattering of electromagnetic waves off of fluids
or other forms of matter, and are found in both classical mechanics (CM) and
quantum mechanics (QM).

The rigged Hilbert space (RHS) approach of the Prigogine school is a method
for solving the equations of a LPS (both CM and QM) consisting in constructing
a complete set of eigenvalues and eigenvectors for the Liouville operator
acting on distribution functions $\rho$.\footnote{These distribution functions
may be understood in terms of the probability density $\rho(\vec{q}_{1}%
,\vec{q}_{2},\vec{q}_{3},...,\vec{p}_{1},\vec{p}_{2},\vec{p}_{3}...,t)$ of
finding a set of molecules (say) with coordinates $\vec{q}_{1},\vec{q}%
_{2},\vec{q}_{3},...$ and momenta $\vec{p}_{1},\vec{p}_{2},\vec{p}_{3}...$ at
time $t$ on the relevant energy surface and are analogous to the
microcanonical distribution.} The construction of such eigenvalues and
eigenfunctions is what Prigogine and colleagues call the `generalized problem
of integration' (Prigogine et al. 1991, p. 4). To be clear about terminology,
finding a transformation that decouples the Hamiltonian in (1) is what is
required to show that the system is completely integrable in the sense
described earlier. Constructing the complete set of eigenvalues and
eigenvectors for a set of equations derived from (1) is what Prigogine and
colleagues refer to as `integrating' or solving the equations of motion.
Although initially motivated in the context of perturbation theory (as
sketched here), the rigged Hilbert space approach is more general in nature
and applicable to any LPS (e.g. most systems in SM, systems involving
interacting fields).

\section{Mathematical Details of the Rigged Hilbert Space Approach}

There are three key elements in the Brussels-Austin method to solving LPS
equations. First, they utilize distribution functions to describe the
dynamics. Second, they adopt extended spaces such as RHS as a mathematical
framework for solving the equations. Third, they introduce an ``appropriate''
time ordering of the dynamical states of the system.

\subsection{The Need for Distributions}

When solutions of the generalized integration problem sketched at the end of
\S2.2 exist, they reduce to classical trajectories for most CM systems and to
state vectors for most QM systems. In the context of a LPS, however, Prigogine
and colleagues argue solutions are not reducible beyond distributions for CM
systems. Examples include systems in kinetic theory, radiation damping and
interacting fields. One important feature of such physical contexts is that
they are characterized by \textit{persistent interactions}. According to
Petrosky and Prigogine, a system's interactions are persistent if there are no
asymptotic states such that the interactions finally cease (1997, pp. 33 and
35). For example in kinetic theory, the molecules of a gas are in constant
interaction with one another because they are undergoing continuous
collisions. This physical situation should be contrasted with the idealized
case of a single neutral particle scattering off a fixed target. In the latter
situation, there is a \textit{transitory interaction} because the particle
undergoes an interaction only in a finite region near the target over a very
short time interval, while the particle spends the majority of its life in the
so-called asymptotic in and out states free of any interactions with the
target. Since interactions never cease for systems with persistent
interactions, the model equations typically will not be completely integrable.

The presence of persistent interactions is one of the features giving rise to
the continuous set of resonances in a LPS. In a gas containing a large number
of particles, these resonances allow for energy to be transferred and leveled
throughout the system. Through persistent interactions and the resulting
resonances, the particles will loose energy and any ordered patterns are
destroyed through \textit{diffusion} (see \S4.2 below).

A further consequence is that the physical dynamics are no longer localized,
but are spread throughout the space occupied by the LPS. For the gas example,
these nonlocal dynamics will take the form of correlations as described in
\S4.2 below. In addition if the number of particles is large enough, then the
degrees of freedom for such a gas of particles will have a continuous spectrum
qualifying it as a LPS. This implies that we should expect the dynamical
description of such systems to be in terms of distributions of particles
rather than in terms of individual particles, because the effects of
long-range and higher-order correlations due to such interactions become at
least as important as the trajectory dynamics. The particles remain coupled to
one another through their interactions resulting in collective effects (\S4.2
below). This type of long-range coupling at least implies that the global or
collective dynamics of the system cannot be accurately represented by
trajectory dynamics alone (see \S4.3 below). As a consequence, Prigogine and
colleagues believe we must view irreversibility as a property of a system that
emerges at the global level which is not derivable from the trajectory
description, meaning that distributions are the natural elements for
representing statistical phenomena rather than trajectories.\footnote{To avoid
a simple confusion (e.g. Bricmont 1995, pp. 165-6), note that singular
distributions such as delta functions \textit{are not} used to represent
probability distributions in the rigged Hilbert space approach.}

\subsection{The Need for RHS}

A RHS is an extended mathematical space first introduced by the Russian
mathematician Gel'fand and his collaborators (Gel'fand and Vilenkin
1964).\footnote{In more recent work Petrosky and Prigogine (1997) have
explored rigging ``Liouville space''--the space of density functions or
density operators--for dynamics. Ord\'{o}\~{n}ez (1998) has demonstrated that
these Liouville spaces can be rigged as a Gel'fand triplet, yielding
semi-group operators and generalized eigenvectors.} Briefly a RHS can be
understood in the following way. Let $\Psi$ be an abstract linear scalar
product space and complete it with respect to two topologies. The first
topology is the standard Hilbert space (HS) topology $\tau_{\mathcal{H}}$%
\begin{equation}
|h|=\sqrt{(h,h)}\text{,}%
\end{equation}
where $h$ is an element of $\Psi$ resulting a HS $\mathcal{H}$. The second
topology $\tau_{\Phi}$ is defined by a countable set of norms%
\begin{equation}
|\phi_{n}|=\sqrt{(\phi,\phi)_{n}}\text{, }n=0,1,2...
\end{equation}
where $\phi$ is also an element of $\Psi$ and the scalar product in (7) is
given by%
\begin{equation}
(\phi,\phi^{\prime})_{n}=(\phi,(\Delta+1)^{n}\phi^{\prime}),n=0,1,2...
\end{equation}
and%
\begin{equation}
\phi_{\gamma}\rightarrow\phi\text{ in }\tau_{\Phi}\text{ iff }|\phi_{\gamma
}-\phi|_{n}\rightarrow0\text{ for every }n\text{,}%
\end{equation}
where $\Delta$ is the Nelson operator $\Delta=\sum X_{i}^{2}$ (Nelson 1959,
587). The $\chi_{i}$ are the generators of an enveloping algebra of
observables for the system in question and they form a basis for a Lie algebra
(Nelson 1959; Bohm \textit{et al}. 1999). For example if we are modeling the
harmonic oscillator, the $X_{i}$ could be the raising and lowering operators
(Bohm 1978, 7-9). Furthermore if the operator $\Delta+1$ is a nuclear operator
then this ensures that $\Phi$ is a nuclear space (Treves 1967, 509-34; Bohm
1967, 276-7). An operator $A$ is nuclear if it is linear, essentially
self-adjoint (Roman 1975, pp. 540-3) and its inverse is
\textit{Hilbert-Schmidt}. The operator $A^{-1}$ is Hilbert Schmidt if $A^{-1}$
$=$ $\sum X_{i}P_{i}$, where the $P_{i}$ are mutually orthogonal projection
operators on a finite dimensional vector space and $\sum a_{i}^{2}<\infty$,
$a_{i}$ denoting the eigenvalues of $A^{-1}$ (Bohm 1967, 273-6). Notice that
the norm (6) is a special case of (7) where \textit{n = 0}.\footnote{There are
many different inequivalent irreducible representations of an enveloping
algebra of a group characterizing a physical system (e.g. the rotation group
has an inequivalent irreducible representation for each value of $j$). They
can be combined in many ways by taking direct products describing combinations
of physical systems. These representations are characterized by the values of
the invariant or Casimir operators of the group. So although the Nelson
operator fully determines the topology of $\Phi$, there is freedom in choosing
the enveloping algebra describing elementary physical systems. Further
restrictions on the choice of function space for a realization of $\Phi$ are
due to the particular characteristics of the physical system being modeled.
This is analogous to the situation for $W^{\ast}$-algebras in the algebraic
approach to QM (Primas 1981 pp. 161-249; Amann and Atmanspacher 1999).}

We obtain a Gel'fand triplet if we complete $\Psi$ with respect to $\tau
_{\Phi}$ to obtain $\Phi$ and with respect to $\tau_{\mathcal{H}}$ to obtain
$\mathcal{H}$. In addition we consider the dual spaces of continuous linear
functionals $\Phi^{\times}$ and $\mathcal{H}^{\times}$ respectively. Since
$\mathcal{H}$ is self dual, we obtain%
\begin{equation}
\Phi\subset\mathcal{H}\subset\Phi^{\times}\text{,}%
\end{equation}
where $\Phi^{\times}$ is characterized by the induced topology $\tau_{\times}%
$. The meaning of the symbol $\subset$ in relation (10) is that every space to
the left of $\subset$ a is a subspace of every space to the right of $\subset$
and every space to the left of $\subset$ is dense in the space to the right of
$\subset$ with respect to the topology of the space to the right of $\subset$
(see Gel'fand and Vilenkin 1964 for more details).

For the Brussels-Austin Group, the chief reason to work in a RHS is the
ability to naturally model unstable physical phenomena such as decay,
scattering and the irreversible approach to equilibrium which is lacking in HS
(e.g., Bishop 2003a). These kinds of time-dependent processes require complex
eigenvalues and generalized eigenfunctions (Gel'fand and Shilov 1967). Such
mathematical quantities are not well-defined in a HS, but are given rigorous
justification in a suitable RHS. In particular the Liouville operator, which
characterizes a LPS's approach to equilibrium, does not have a complete set of
eigenvalues and eigenfunctions in a HS. Recently the Brussels-Austin Group has
demonstrated that a complete set of eigenvalues and eigenvectors for this
important operator can be defined and calculated for several chaotic models in
extended spaces (Antoniou and Tasaki 1992 and 1993; Hasegawa and Shapir 1992;
Hasegawa and Driebe 1993). An additional motivation for switching to a RHS is
that the equations of motion defined on a HS are time-symmetric.
Time-asymmetric equations may be defined and solved in a RHS making the latter
type of space a natural choice for modeling intrinsic irreversible processes
(irreversibility without explicit reference to an environment; see Part I).
Intrinsic irreversibility is of prime interest to the Brussels-Austin Group
because these types of irreversible processes are related to intrinsic arrows
of time in physics (i.e. arrows of time which are independent of human
intervention or approximation).

\subsection{Semigroup Operators in RHS and Irreversibility}

One of the important features of RHS is that evolution operators are often
elements of semigroups rather than groups, so that irreversible behavior can
be appropriately modeled. The case of simple scattering is a good example for
illustrating the concepts. An idealized version of a scattering experiment is
sketched in Figure 1. There is a preparation apparatus which prepares
particles in a particular state (energy, angular momentum, etc.). The
particles are emitted at a target (assumed to be fixed in this analysis). The
free particle Hamiltonian in (1) is $H_{0}$ while the potential in the
interaction region surrounding the scattering center is given by $V$. After
the interaction with the target, the detector registers the particle measuring
quantities such as the angle of scattering relative to the initial direction
of the particle as emitted from the accelerator or the energy of the particle
after the scattering event.%

\begin{figure}
[ptb]
\begin{center}
\includegraphics[
natheight=2.068600in,
natwidth=5.344500in,
height=2.1075in,
width=5.4008in
]%
{scattering.wpg}%
\caption{An idealized scattering experiment.}%
\end{center}
\end{figure}

Each interaction involves a resonance which can be described as%
\begin{equation}
|E^{\pm}>=\left(  1+\frac{1}{E-H\pm i\varepsilon}V\right)  |E>\text{,}%
\end{equation}
a Lippmann-Schwinger-type equation for the evolution of the energy eigenstates
as they pass through the scattering region. Whenever the operator on the right
hand side of (11) applied to the energy eigenstate $|E>$ goes to infinity, we
have a resonance. According to the Brussels-Austin Group, if, given a
sufficiently large number of interacting particles, the number of resonances
in a system is sufficiently large, then the system will evolve from a highly
ordered state to a completely randomized or equilibrium state. This evolution
is intrinsically irreversible, due to the internal dynamics of the system.

The intrinsic irreversibility of LPS models must be described by semigroups.
This necessitates leaving the HS framework and working in a broader
mathematical space such as a RHS which Antoniou and Prigogine (1993) adopt in
their analysis of the Friedrich's model for scattering. In the Gel'fand
triplet $\Phi\subset\mathcal{H}\subset\Phi^{\times}$, $\Phi^{\times}$ is the
space of particle distribution functions. Furthermore Antoniou and Prigogine
adopt the following time ordering condition: any excitations or preparations
are to be interpreted as events taking place before $t$ = 0 while any
de-excitations or detections are to be interpreted as events taking place
after $t$ = 0 (1993, pp. 445 and 455).

At the point in the analysis of the scattering experiment where choices have
to be made regarding how to interpret the directions of integration for the
analytic functions involved in the upper and lower complex half-planes, they
choose the following interpretations (1993, pp. 454-5): excitations are
identified as taking place \textit{before }$t=0$ (taken to be represented as
extensions from the lower to the upper half-plane), while de-excitations are
identified as taking place after\textit{\ }$t=0$ (taken to be represented as
extensions from the upper to the lower half-plane). So the time-ordering rule
is applied with respect to the choice of how to deform the contours in the
complex plane with respect to the choice of direction of integration along the
contours. Proceeding in this fashion Antoniou and Prigogine derive concrete
realizations for the space $\Phi$ involving Hardy class function spaces (1993
pp. 457-9; see also Bishop 2003a and 2003b).

Antoniou and Prigogine discuss two semigroups of evolution operators. The
first is $U^{\dagger}(t)=e^{-iHt}$, initially defined on $\mathcal{H}$ for
$-\infty<t<\infty$, extended to $\Phi^{\times}$. It is continuous and complete
in the topology $\tau_{\times}$ of $\Phi^{\times}$, valid for $t\geq0$ and
they identify its temporal direction as carrying states into the forward
direction of time. This operator describes evolution reaching equilibrium in
the future. The second operator is $U^{\dagger}(t)$ extended to $\Phi^{\times
}$, continuous and complete in the topology $\tau_{\times}$, but valid for
$t\leq0$.\footnote{The requirements of continuity and completeness force the
unitary group extended to $\Phi^{\times}$ to be restricted to the separate
time ranges $t\leq0$ and $t\geq0$ (Bohm and Gadella 1989, pp. 35-119).} They
identify the temporal direction of this latter operator as carrying states
into the \textit{backward direction} of time ($-t$ increasing), so this
operator describes evolution reaching equilibrium in the past. Since no
physical systems are ever observed evolving to equilibrium from the future
into the past, they \textit{select} $U^{\dagger}(t)$ extended to $\Phi
^{\times}$ for $t\geq0$ as the physically relevant semigroup of evolution
operators for modeling statistical mechanical systems. This selection is taken
to be an expression of the second law of thermodynamics based on our empirical
observations (Antoniou and Prigogine 1993, p. 461).

The approach sketched in this section for the case of transient scattering can
be extended to the case where the interactions are continuous and persistent,
yielding similar results (Petrosky and Prigogine 1996 and 1997).

\section{Discussion of the RHS Approach}

The Brussels-Austin Group's RHS approach has yielded solutions (mostly
numerical) to nonequilibrium statistical mechanical system equations. Based on
these solutions and the insights gained from the new approach, Prigogine and
coworkers make a number of important claims needing detailed discussion.

\subsection{Thermodynamic Arrow of Time}

One of the claimed virtues of the approach is that it provides an explanation
for the thermodynamic arrow of time (the law of increasing entropy defined
entropy close to equilibrium). This has been one of the central goals of
Prigogine since he began his work in SM. One feature that both the earlier
similarity transformation approach (discussed in Part I) and the RHS approach
share in this quest is a kind of vacillation between seeking an explanation of
the thermodynamic arrow in the dynamics of the physical system, and taking the
empirically observed direction of the arrow as a fundamental principle.

In the RHS approach, the types of mechanisms to which the Brussels-Austin
Group appeals for explaining the thermodynamic arrow are diffusion, the growth
of correlations and collective effects, all of which are generated by
Poincar\'{e} resonances (Antoniou and Prigogine 1993; Petrosky and Prigogine
1996 and 1997). The extension of the description of a LPS with their
Poincar\'{e} resonances, persistent interactions and chaotic dynamics to
Gel'fand triplet spaces allows the eigenvector equations to be solved. In the
course of analyzing these solutions, characteristically there are two
semigroups that emerge as sketched in \S3.3. At this point in the analysis,
one semigroup is selected because it represents systems approaching
equilibrium in the temporal direction of the future, while the other semigroup
is disregarded because it describes systems approaching equilibrium in the
temporal direction of the past which is never observed and, therefore, deemed
to be unphysical (Antoniou and Prigogine 1993, p. 461; Petrosky and Prigogine
1996, p. 453 and 1997, p. 13). By making this latter appeal to observations,
the Brussels-Austin Group is appealing to the very facts they seek to explain
\textit{via} the dynamics of the physical system.

The model equations alone do not uniquely determine which semigroup is the
appropriate one, so some kind of appeal to physical considerations is needed.
As discussed in \S3.3 above, the Brussels-Austin Group does make an appeal to
a criterion for choosing a temporal ordering: any excitations are to be
interpreted as events taking place before $t=0$ while any de-excitations are
to be interpreted as events taking place after $t=0$. While there is a clear
ordering of time from excitation to de-excitation, the criterion invoked still
ultimately rests upon our observations that a system is excited before it
undergoes de-excitation. The \textit{physical} reason why the thermodynamic
arrow runs from the past toward the future is still undiscovered in the RHS
approach, though the approach gives us the mathematical tools to explore and
describe the arrow precisely.

\subsection{Correlation Dynamics}

The RHS approach highlights the role of nonlocal and collective effects due to
long-range correlations that introduce new dynamics in the probabilistic
description that are typically ignored in the trajectory description of a
LPS.\footnote{Prigogine (1962, 138-95) introduced a simplified version of
correlation dynamics and George (1973a) developed the idea in the direction
indicated in this section.} The term ``collective effects'' is used to
describe the behavior of an aggregate of particles coupled together in some
fashion that is distinct from the behavior of individual particles. Collective
effects can arise from long-range forces such as electromagnetism, gravity or
from spatial correlations caused by interactions.

Spatial correlations play an important role in the temporal ordering of the
dynamics of SM systems. In atomic or molecular gases, collective effects are
due to collisions. Consider the idealized textbook situation, where we start
with an isolated gas of $N$ particles in a volume $V$ that have yet to
interact with one another. If the initial distribution of the particles is
homogeneous and isotropic, then the particles are equally likely to be at any
point $\vec{r}$ in $V$.\footnote{Of course, in this idealized example the
assumption of equiprobability of states is reasonable. In a LPS, by contrast,
interactions are persistent, so this assumption cannot be made.} This result
holds for each individual particle under the condition that the positions of
the other particles are arbitrary. In a typical gas or liquid, this latter
condition is not fulfilled in general, however. Consider two particles at a
time in our gas. Given the position of one particle, different positions of
the second particle are not equally likely to obtain; namely, the second
particle cannot occupy the position of the first particle. Due to
interparticle interactions and the symmetry properties of the state vectors,
different values of the relative position $(\vec{r}_{2}-\vec{r}_{1})$ between
our two test particles in the entire gas do not appear with equal likelihood.
This feature is known as a \textit{spatial correlation} between the
simultaneous positions $\vec{r}_{1}$ and $\vec{r}_{2}$ of the two particles.

\qquad In a plasma, for example, where the gas is composed of charged
particles, spatial correlations are the tendencies of unlike charges to
cluster together and the tendencies of like charges to repel each other. The
simultaneous positions of the particles in the plasma are not all equally
likely. It turns out that there is a simple relationship between the spatial
integral of the correlation function representing spatial correlation and the
mean square fluctuation of the density of the gas particles (Pathria 1972,
447-50), meaning the spatial distribution of the particles is influenced by
the presence of such correlations. In addition these correlations are directly
dependent on the density of particles in the gas. As the density decreases,
such collective effects disappear because the \textit{mean free path} of the
particle, a measure of the likelihood of a collision during a given distance
traveled, becomes comparable to $V$. This means collision events will be very
rare and correlations will be kept to a minimum when the mean free path is large.

Collisions are frequent in dense gases and the spatial correlations induced by
collisions couple each particle with many other particles in the gas. It is
this coupling due to correlations that leads to collective behavior
responsible for gas particles being collected into coherent structures rather
than being uniformly spread throughout the volume. Examples would be
turbulence and shock waves.

To see how these correlations develop, start with the particles in the gas
before they have interacted with each other. As they begin colliding, the
first interactions set up binary correlations between particles. As the
interactions persist, ternary correlations begin to appear. The process will
continue by establishing quaternary correlations and so on through N-ary
correlations as more and more particles become involved others through
collisions. The progression from lower order correlations (which appear first)
to higher order correlations (which appear later) corresponds to a natural
temporal ordering for the evolution of the states of the gas. Correlations and
other collective effects can rival or exceed the role of individual particle
trajectories and be masked by a dynamical description that treats trajectories
as its basic explanatory element.

For example the electromagnetic force is a long-range force. It is the
dominant force in many situations in a plasma, so the behavior of a plasma is
not reducible to the dynamics of the trajectories of the individual particles
alone. In the case of a plasma, the energy of the plasma is affected by the
presence of correlations, such that one of the differences between the energy
of a plasma and that of an ideal gas (noninteracting particles) is given by a
correction term due to correlation effects (Krall and Trivelpiece 1986, 63-5).
Not only do these effects interact with the electromagnetic fields of the
plasma itself, but they also generate new electromagnetic fields that react
back on the plasma leading to very complex dynamics.

Among the physical mechanisms playing a role in LPS, correlations appear to
play a crucial role in irreversibility. As was apparent in the earlier
similarity transformation approach, the progression of correlations suggests a
natural direction for the thermodynamic arrow (George 1973a). But this is not
simply another way of saying that entropy increases for such systems because
in an open system the order of correlations may continue to grow while the
measure of disorder in the system may remain constant or decrease. So
correlations are not the complete explanation for the thermodynamic arrow of time.

Long-range correlations are another effect in the dynamics of correlations
that become apparent in RHS (discussed in its earliest form in Prigogine 1962
and George 1973a). As gas particles begin to interact, correlations develop
among the particles due to interactions (recall that in a LPS these
interactions are associated with resonances). Along with the growing order of
correlations, long-range correlations develop as particles interact with one
another and then separate over long distances while carrying the ``memory'' of
their prior interactions (correlations) with them to other parts of the gas.
Over short time scales, the growing order of correlations appears to be the
more dominant of the two effects. As time goes on, the long-range correlations
due to resonances are built up so that collective effects become influential.
These long-range correlations are associated with nonequilibrium modes of
energy transfer (Petrosky and Prigogine 1996, p. 468).

Over longer time-scales, another very interesting phenomenon occurs.
Equilibrium short-range binary correlations remain finite, but nonzero around
each particle. In turn ternary nonequilibrium correlations are built up among
particles in a small region. These correlations diffuse throughout the system,
leaving the equilibrium correlations, while quartinary nonequilibrium
correlations are built up among the local particles. These correlations
diffuse throughout the system while quintinary nonequilibrium correlations
build up and so forth. As time continues the variously ordered nonequilibrium
correlations can propagate over large distances due to diffusion so that the
corresponding information is transferred globally among the particles of the
gas. The end result is a ``sea'' of multiple incoherent correlations (Petrosky
and Prigogine 1996, p. 468). This effect provides a natural temporal direction
for the flow of entropy and is revealed in the types of complex spectral
representations of the statistical evolution operators made possible by
working in an RHS.

In this sense one might argue that as the order of correlations increases, as
long-range correlations grow and as higher-order nonequilibrium correlations
propagate throughout the gas, the effects of individual trajectories on the
global dynamics of the gas become less important relative to the effects of
the dynamics of correlations. This does not mean that particles lack
trajectories and positions in state space as these types of interaction events
are parasitic on these concepts (e.g. mean free path between collisions). In
my view correlations and collective effects make the significant contributions
to the global dynamics while the effects of trajectories play a role only
locally (see below).\footnote{Of course I have used idealized examples in this
section in the sense that we imagined starting with a gas of noninteracting
particles and then ``turning on'' the interactions. Recall that interactions
are persistent in a LPS so there is never a time in such systems when the
microscopic dynamics can be characterized by smooth, smooth trajectories.}

One might object that the dynamics of correlations can somehow be reversed
even though the probability of the right kinds of reversals to run the whole
evolution backwards (like a film in reverse) is extremely small. If true, then
the situation is still the same as in standard thermodynamics where the
increase in entropy in systems is viewed as being reversible though the
probability is vanishingly small.

The Brussels-Austin response to such an objection for open systems has been
given in \S4 of Part I. For closed systems they have shown that as the
dynamics of correlations continue, an ``entropy barrier'' against inversion
develops. This barrier can be defined as the value of the \textsl{H}%
-function--a thermodynamic function related to the entropy, which does not
require coarse graining or the invocation of an environment in the
Brussels-Austin approach--after such an inversion minus its value before such
an inversion. This difference increases exponentially with time, so the longer
the LPS evolves, the higher the barrier to inversion. Essentially this means
that the energy requirements to invert the system of particles increases very
rapidly with time. As the model approaches equilibrium, this energy barrier
diverges, hence, there is no physical way of ``going back'' in the
anti-thermodynamic direction (Petrosky and Prigogine 1996, pp. 468-9 and 494-5).

\subsection{``Collapse of Trajectories''}

In the similarity transformation approach (Part I), Prigogine and
collaborators put forward several arguments to the effect that smooth (i.e.,
everywhere differentiable), deterministic trajectories do not exist for
unstable statistical mechanical systems. These arguments were fundamentally
flawed in similar ways in that epistemological claims were treated as
ontological claims. In the new approach, this bias against such smooth
trajectories and the dynamics derivable from trajectories resurfaces in a
different form that clarifies the Brussels-Austin attitude toward trajectories.

It is well known that in the traditional description, the trajectory of a
point particle free of any external forces can be represented mathematically
as a superposition of ``plane waves'' by taking the position of the particle
and applying a Fourier transform from $(q,p)$ space to $(k,p)$ space. In this
latter space, a trajectory is a coherent superposition of plane waves and this
superposition is modeled by a Dirac delta function. For a particle undergoing
free motion, this distribution function is a solution to the equation of
motion, has unchanging width and is everywhere differentiable throughout its
deterministic evolution (``smooth'' trajectory).

For a finite number of particles with normalizable distributions, the
trajectory description in $(k,p)$ space and the Brussels-Austin probabilistic
description agree.\footnote{Some critics, such as Bricmont (1995, pp. 165 and
175), have overlooked the way in which the RHS approach reduces to standard SM
approaches for small numbers of particles when LPS conditions are not
fulfilled.} In the thermodynamic limit, however, Prigogine and coworkers argue
that resonances destroy smooth trajectories in the following way. In the
thermodynamic limit, the Dirac delta function describing the trajectories of
particles at $t=0$, once evolution begins, immediately begins spreading
throughout a subspace of $(k,p)$ space under the action of resonances, though
maintaining a delta function singularity\footnote{The significance of the
delta function singularity appears to be more mathematical than physical.
Mathematically it means that so-called reduced distribution functions--where
the distribution function refers to a subset $s$ of the total number of
particles in the system--exists in the thermodynamic limit, but such
distribution functions almost always exist for molecules under most realistic
forces. Reduced distributions were introduced into nonequilibrium contexts by
(Brout and Prigogine 1956; Prigogine and Balescu 1959).} (Petrosky and
Prigogine 1996, pp. 479-481 and 1997, pp. 35-37). The trajectories are no
longer representable as delta functions, but by broader kinds of distribution
functions. Petrosky and Prigogine unfortunately described this phenomenon as
the ``collapse of trajectories'', but all they really mean is that a different
notion of trajectory is required in a LPS.

In $(q,p)$ space, this implies that there are no longer any smooth (everywhere
differentiable) trajectories, but, rather, trajectories exhibiting
\textit{Brownian motion}. A simple way to see this is to return to our
idealized gas example. As before, assume initially that the particles have not
interacted with each other. Prior to any collisions, the motion of the
particles can be characterized by smooth trajectories. As they begin
interacting, the particle trajectories become piece-wise continuous as
instantaneous discontinuities arise associated with each collision. Continuous
interactions of this type would then prevent trajectories from being
everywhere differentiable, resulting in particles exhibiting Brownian
trajectories rather than smooth ones, but this in no way implies that there
are no trajectories whatsoever.

Consider the special case of a single smooth trajectory represented as%
\begin{equation}
\gamma(p,q)=\prod_{i=1}^{N}\delta(\vec{q}_{i}-\vec{q}_{i}^{0})\delta(\vec
{p}_{i}-\vec{p}_{i}^{0})
\end{equation}
in a LPS model where the superscript $0$ indicates the contribution from the
unperturbed Hamiltonian. To first order the time evolution of the momentum for
the component $i=1$ is giving by%
\begin{equation}
\vec{p}_{1}(t)=\vec{p}_{1}^{0}+\lim_{\Omega\rightarrow\infty}\frac{\lambda
}{\Omega}\sum_{k}\sum_{n=2}^{N}(-\vec{k})\frac{V_{k}}{\vec{k}\cdot(\vec{v}%
_{1}^{0}-\vec{v}_{n}^{0})-i\varepsilon}\left(  e^{-i\vec{k}\cdot(\vec{v}%
_{1}^{0}-\vec{v}_{n}^{0})t}-1\right)  e^{-i\vec{k}\cdot(\vec{q}_{1}^{0}%
-\vec{q}_{n}^{0})}\text{,}%
\end{equation}
where $\Omega$ is the volume, $\vec{k}$ is the wave vector, $\vec{v}_{1}$ is
the velocity vector of particle 1, $\vec{v}_{n}$ is the velocity vector of
particle $n$, and $\varepsilon$ is an infinitesimal positive constant. The
first term represents the contribution from the unperturbed Hamiltonian and
the second term represents contributions from the interactions. If $N$ is
finite, (13) becomes%
\begin{equation}
\vec{p}_{1}(t)=\vec{p}_{1}^{0}+\lambda\sum_{n=2}^{N}\int d\vec{k}\frac{V_{k}%
}{\vec{k}\cdot(\vec{v}_{1}^{0}-\vec{v}_{n}^{0})-i\varepsilon}\vec{k}%
e^{-i\vec{k}\cdot(\vec{q}_{1}^{0}-\vec{q}_{n}^{0})}+O(\lambda^{2})\text{,}%
\end{equation}
in the limit $t\rightarrow\infty$ because the pole at $\vec{k}\cdot(\vec
{v}_{1}^{0}-\vec{v}_{n}^{0})=i\varepsilon$ vanishes as $\Omega\rightarrow
\infty$, the LPS condition. According to (14) the value of the momentum to
first order asymptotically approaches a constant and the time dependence drops
out. Note that in the limit $|\vec{q}_{1}^{0}-\vec{q}_{n}^{0}|$ $\rightarrow
\infty$, the interactions from particles $n$ remains finite even if such
interactions are short-ranged due to resonances, so that long-range
correlations are built up. In the thermodynamic limit, (13) generally diverges
and Petrosky and Prigogine conclude that point distributions such as (12)
representing trajectories are not physically admissible and, therefore, smooth
trajectories are inconsistent with the thermodynamic limit in a LPS (1996, p.
480). Only singular nonlocal distributions appear to be consistent with the
thermodynamic limit and such distributions lie outside of HS (Petrosky and
Prigogine 1996, pp. 479-81).

These results are related to the nonlocal nature of the collective effects of
the entire distribution described in \S4.2 above. If any arbitrary finite
number of particles were selected within the system and treated in isolation,
all nonlocal diffusion and correlation effects become negligible and we are
left with the standard description and results in terms of trajectories
(however, these trajectories would not necessarily be everywhere differentiable).

In more realistic situations, the nonexistence of smooth trajectories leads
directly to the Brussels-Austin claim that a LPS exhibits behavior that
\textit{cannot be derived from trajectory dynamics}. Such effects include the
breaking of time symmetry (i.e., the appearance of semigroups of operators
governing the evolution instead of groups), diffusion and nonlocal
correlations. Prigogine and coworkers refer to these effects as
``non-Newtonian'' to emphasize the fact that the trajectory description is
inadequate to account for them. The existence of collision operators such as
the Fokker-Planck operator is only a necessary condition for irreversibility
and other ``non-Newtonian'' effects. Particular types of distributions (namely
singular distributions) must also be present in order to have sufficient
conditions for such behavior. The class of singular distribution functions is
quite broad and applicable to many ordinary situations in SM (the canonical
distribution is an example; see also Prigogine 1962 and 1997). Petrosky and
Prigogine have carried out algebraic and computer modeling to demonstrate that
the trajectory and distribution descriptions yield different results for LPS
(e.g. 1993, 1994 and 1996).

I believe the appropriate way to understand this new approach with its
``non-Newtonian'' effects is to agree with them that distribution descriptions
cannot be reduced to point-wise descriptions. However, both descriptions
should be viewed as valid within their domains. The trajectory description is
valid for local regions of a LPS, where there are relatively few particles, so
that trajectory dynamics is the dominant feature (the trajectories may be
either smooth and exact, or exhibit random walks). Interactions take place
among particles at this local level and to the extent that we can ignore
higher-order and long-range correlations, trajectory and distribution
descriptions agree in their account of physical behavior as was noted earlier.

Where my interpretation of the Brussels-Austin work differs from their own is
when the conditions for a LPS are met (large number of particles, continuous
frequencies, etc.). I agree that in examining the global evolution of LPS,
higher-order correlations and collective effects due to long-range, persistent
interactions are the dominant features, which are not reducible to trajectory
dynamics alone. Trajectories are not irrelevant, however, because such
features as correlations and collective effects presuppose particle positions
and trajectories. For example, collective effects in ordinary gases do not
disconfirm the existence of trajectories, though the effects of correlations
can rival or exceed the effects of individual particle trajectories and be
masked by a dynamical description that treats trajectories as the sole
explanatory element. Note that (14) does not imply smooth point trajectories
are immediately expunged from a LPS. Physically smooth trajectories are
converted into random walks due to the persistent interactions and the
long-range higher-order correlations that diffuse throughout the system over
time. As described above, resonances, collisions and correlations are closely
related to long-range correlations and collective effects, behavioral features
of unstable systems for which the trajectory description alone cannot
adequately account. For LPS models the whole is more than the sum of its
parts. Particle trajectories are necessary for global distributions to exist,
but are insufficient for determining how such global distributions evolve in
time. The thermodynamic paradox might be dissolved because (1) the
time-symmetric behavior of the trajectory dynamics contributes nothing more to
the global evolution of the SM system than the necessary conditions for the
existence of such a system and (2) in a LPS trajectories exhibit Brownian
motion and correlation dynamics dominate the macroscopic dynamics.
Thermodynamic behavior is, then, an emergent global phenomenon possessing a
temporal direction.

My interpretation suggests a way to reduce the tension in their view between
operationalism with respect to trajectories and realism with respect to
distributions (see Part I), where the Brownian trajectories of the system give
the \textit{necessary} conditions for the existence of the distribution $\rho
$, but \textit{not sufficient} conditions for its evolution. In my judgement
the new approach the Brussels-Austin Group has been exploring illuminates some
of the underlying physical mechanisms of thermodynamic behavior. Focusing on
the growth and dynamics of correlations and collective effects are important
physical insights which have advanced our understanding of thermodynamics
processes. And by employing extended mathematical structures such as RHS, they
have developed powerful tools for describing such processes which will
doubtless lead to further insights.

As a last comment, I should point out that this RHS approach does not
represent a kind of coarse-graining approach, at least as normally understood.
Emphasis shifts away from trajectories because they are only a part of the
story of the behavior of a LPS (coarse-grained accounts typically assume that
trajectory dynamics is the whole story, but that complete descriptions at the
trajectory level are computationally intractable). And, as in the similarity
transformation approach, the RHS approach distinguishes between manifolds of
stable and unstable motions (in contrast to typical coarse-grained accounts).
Furthermore, if the global behavior of a LPS is not only emergent, but also
constrains the motion of individual particles (say by restricting the modes of
energy transfer), then an appropriate mathematical description should be able
to describe this kind of feedback between levels in a system. The RHS approach
can describe such feedback effects, whereas coarse-grained accounts cannot
because they deal with only one level of a given system. Finally, whether
trajectories that are not everywhere continuous nor everywhere differentiable
are deterministic or not is an open question in the RHS approach, as I discuss
in the next section (coarse-grained accounts typically assume trajectories are
deterministic, though usually no explicit assumptions are made regarding the
trajectories' continuity and differentiablility).

\section{Possibility Rather than Certainty?}

Prigogine's provocatively titled book, \textit{The End of Certainty} (1997),
sums up one of arguably the most important and far reaching consequences of
the Brussels-Austin Group's work: Namely, that the certainty of the
deterministic, time-symmetric trajectory description is not applicable to the
global dynamics of a LPS. Instead only a statistical description of
probability densities remains. In conventional CM and SM models, particle
positions and trajectories are treated as the fundamental ontological entities
determining the dynamical evolution of the system. In the Brussels-Austin view
this is no longer the case for LPS models. The fundamental ontological feature
for these models are the probability distributions, i.e., the large-scale
arrangements of the particles themselves. To reformulate the laws of classical
dynamics along the statistical lines suggested by Prigogine and co-workers
leads to the conclusion that such laws now `express ``possibilities'' and no
more ``certainties''' (Petrosky and Prigogine 1997, p. 1).

Where there are relatively few numbers of particles, the Brussels-Austin
Group's approach to dynamics reduces to the standard results of CM, so the
trajectory picture with its deterministic and time-reversible character is
preserved as a limiting case. In non-LPS cases, the RHS approach recovers the
usual results of SM (e.g. Fokker-Planck equation, Boltzmann equation,
non-Markovian master equations). It is in cases where the LPS criteria apply
that probability becomes the fundamental notion, irreducible to the trajectory
description. Systems must be treated as wholes. If any subset of the total
number of particles $N$ is treated by itself all the ``non-Newtonian'' effects
disappear and the conventional descriptions are recovered. It is in this sense
that Prigogine believes, `What is now emerging is an ``intermediate''
description that lies somewhere between the two alienating images of a
deterministic world and an arbitrary world of pure chance...[T]he new laws of
nature deal with the possibility of events, but do not reduce these events to
deductible, predictable consequences' (Prigogine 1997, p. 189).

The nature of this possibility supposedly represents a new conception which
remains to be clarified, however. It is clearly not the kind of irreducible
indeterminism described in von Neumann collapse, where some sort of collapse
from multiple possibilities to a single actuality is envisioned. As Prigogine
and colleagues describe it, their probabilistic formulation of physics is also
to be distinguished from the type of chaotic dynamics, where the underlying
dynamics is deterministic, but the outcomes of the system are not predictable.
The latter is \textit{epistemically} indeterminable but not \textit{ontically}
indeterministic.\footnote{Understanding what it means for a system or a
description to be ontically indeterministic is by no means straightforward
(e.g. Bishop 2002).} Instead the dynamics envisioned by Prigogine and his
colleagues involve an interplay between unitary reversible processes and
irreversible processes. The LPS are important examples of dynamical systems
which show this kind of interplay and are, therefore, intrinsically probabilistic.

\qquad But the relationship of this probabilistic evolution to deterministic
dynamics remains unclear and requires attention because under some conditions
the dynamics of probability distributions can be ``embedded'' into completely
deterministic dynamics and Markov processes can almost always be ``embedded''
into deterministic Kolmogorov processes (Antoniou and Gustafson 1993;
Gustafson 1997, pp. 55-76). This leaves open the possibility that there is no
\textit{significant fundamental difference between this new conception of
probabilistic evolution and the conventional conception of deterministic
evolution}, or so one could plausibly argue.\footnote{I should point out that
although there may exist theorems showing that given any Markov process, that
process can be embedded in a larger deterministic Kolmogorov process, the
general result does not necessarily mean that the given Markov process is
deterministic. Whether or not a given Markov process is deterministic or not
is an ontological rather than a mathematical question. It should also be
clear, however, that simply characterizing the probability densities via
Kolmogorov measures is insufficient because this cannot settle the ontological
nature of the probability.}

Though more needs to be said regarding the notion of probabilistic dynamics
they are working out, it must be internally generated by the dynamics of the
system (e.g. \textit{via} correlation dynamics) rather than imposed from the
outside by observers, measuring apparatuses or the environment. I do not take
it that this need for more clarification is a serious weakness of their
program. On the contrary, I look at the situation as analogous to the early
days of quantum theory where many concepts (indeterminacy being one of them)
were very hazy at the start inviting serious reflection and exploration.

The RHS formalism gives us a unified description of dynamics and
thermodynamics within a statistical framework and a consistent, rigorous
description of irreversible processes. The mathematical developments are
indeed impressive, including new results regarding the theory of complex
spectral representations of operators. Furthermore this framework is powerful
enough to allow a unification between CM and QM (Prigogine et al. 1991;
Petrosky, Prigogine and Tasaki 1991; Petrosky and Prigogine 1994). However,
the promise of the recent Brussels-Austin work must be balanced against two
important open questions: (1) What is the physical and mathematical status of
the past-directed $t\leq0$ semigroup (\S3.3) and (2) What is the precise
nature of the probability lying at the heart of an LPS? Answering these two
questions holds the key to their being able to offer an explanation for the
thermodynamic arrow of time and for their developing a notion of indeterminism
that is different in kind from that discussed in conventional QM developments
that would be truly revolutionary.

As things stand, the Brussels-Austin Group has given us a powerful descriptive
tool for irreversible processes, and nonlinear dynamics more generally, but
they have not given us an explanation for the origination of the
irreversibility we observe in our world. One might object that the RHS
approach is ultimately only of mathematical interest since there is nothing
philosophically interesting given the current state of the above open
questions. This response is too quick, however. These open questions can also
be viewed as opportunities for exploration of the underlying concepts of the
approach in order to attempt to answer these questions. For example, by
adopting a different arrow of time in the context of scattering in a RHS
formulation of QM, one can show that the $t\leq0$ semigroup is also future
oriented (this time arrow is, however, highly operational in character and not
generally applicable outside of laboratory contexts; for discussion, see
Bishop 2003a and 2003b). So interesting conceptual questions are raised by the
Brussels-Austin work. Besides, even if questions (1) and (2) should ultimately
be answered in a way that closes off this avenue for nonequilibrium SM, that
information is also valuable to philosophers.

\textit{Acknowledgments}--This essay was considerably strengthened from
discussions with and comments by Ioannis Antoniou, Harald Atmanspacher, Jeremy
Butterfield, Dean Driebe, Fred Kronz, Tomio Petrosky, Michael Redhead, Michael
Silberstein and numerous anonymous referees. I take full responsibility for
remaining weaknesses.

\bigskip

\textbf{References}

\bigskip

Amann, A. and Atmanspacher, H. (1999) `$C^{\ast}$- and $W^{\ast}$-Algebras of
Observables, Their Interpretations, and the Problem of Measurement', in H.
Atmanspacher, A. Amann and H. M\"{u}ller-Herold (eds), \textit{On Quanta,
Mind, and Matter: Hans Primas in Context} (Dordrecht, The Netherlands: Kluwer
Academic Publishers).

Antoniou, I. and Gustafson, K. (1993) `From Probabilistic Descriptions to
Deterministic Dynamics'', \textit{Physica A} \textbf{197}: 153-66.

Antoniou, I., Gustafson, K. and Suchanecki, Z. (1998) `From Stochastic
Semigroups to Chaotic Dynamics', in A. Bohm, H-D. Doebner and P. Kielanowski
(eds.), \textit{Irreversibility and Causality: Semigroups and Rigged Hilbert
Spaces} (Berlin: Springer-Verlag).

Antoniou, I. and Prigogine, I. (1993) `Intrinsic Irreversibility and
Integrability of Dynamics', \textit{Physica A} \textbf{192}: 443-464.

Antoniou, I. and Tasaki, S. (1992) `Generalized Spectral Decomposition of the
$\beta$-adic Baker's Transformation and Intrinsic Irreversibility'
\textit{Physica A} \textbf{190}: 303-329.

\_\_\_\_\_\_\_\_\_\_\_\_\_\_\_\_ (1993) `Spectral Decomposition of the Renyi
Map', \textit{Journal of Physics} \textbf{A26}: 73-94.

Bishop, R. (2002) `Deterministic and Indeterministic Descriptions', in H.
Atmanspacher and R. Bishop (eds.), \textit{Between Chance and Choice:
Interdisciplinary Perspectives on Determinism} (Thorverton: Imprint Academic),
in press.

\_\_\_\_\_\_\_\_ (2003a) `The Arrow of Time in Rigged Hilbert Space Quantum
Mechanics', \textit{International Journal of Theoretical Physics}, in press.

\_\_\_\_\_\_\_\_ (2003b) `Quantum Time Arrows, Semigroups and Time-Reversal in
Scattering', \textit{International Journal of Theoretical Physics}, accepted.

Bohm, A. (1967) `Rigged Hilbert Space and Mathematical Description of Physical
Systems', in W. Brittin, A. Barut and M. Guenin (eds), \textit{Lectures in
Theoretical Physics Vol IX A: Mathematical Methods of Theoretical Physics}
(New York: Gordon and Breach Science Publishers, Inc), pp. 255-317.

\_\_\_\_\_\_\_ (1978) \textit{The Rigged Hilbert Space and Quantum Mechanics}
(Berlin: Springer-Verlag).

Bohm, A. and Gadella, M. (1989) \textit{Dirac Kets, Gamow Vectors, and
Gel'fand Triplets, Lecture Notes in Physics, vol. 348} (Berlin: Springer-Verlag).

Braunss, G. (1984) `On the Construction of State Spaces for Classical
Dynamical Systems with a Time-dependent Hamiltonian Function', \textit{Journal
of Mathematical Physics} \textbf{25}, 266-70.

Bricmont, J. (1995) `Science of Chaos or Chaos in Science?' \textit{Physicalia
Magazine} \textbf{17}: 159-208.

Brout, R. and Prigogine, I. (1956) `Statistical Mechanics of Irreversible
Processes, Part VII: A General Theory of Weakly Coupled Systems',
\textit{Physica} \textbf{22}: 621-36.

Gel'fand, I. and Shilov, G. (1967) \textit{Generalized Functions Volume 3:
Theory of Differential Equations}, Meinhard E. Mayer tr. (New York: Academic Press).

Gel'fand, I. and Vilenkin, N. (1964) \textit{Generalized Functions Volume 4:
Applications of Harmonic Analysis}, Amiel Feinstein tr. (New York: Academic Press).

George, C. (1973a) `Subdynamics and Correlations', \textit{Physica}
\textbf{65}, 277-302.

\_\_\_\_\_\_\_\_ (1973) \textit{Lectures in Statistical Physics II, Lecture
Notes in Physics} (Berlin: Springer-Verlag).

Gustafson, K. (1997) \textit{Lectures on Computational Fluid Dynamics,
Mathematical Physics, and Linear Algebra} (Singapore: World Scientific).

Hasegawa, H. and Driebe, D. (1993) `Spectral Determination and Physical
Conditions for a class of chaotic Piecewise-Linear Maps' \textit{Physics
Letters} \textbf{A176}: 193-201.

Hasegawa, H. and Shapir, W. (1992) `Unitarity and Irreversibility in Chaotic
Systems', \textit{Physical Review} \textbf{A46}: 7401-7423.

Hilborn, R. (1994) \textit{Chaos and Nonlinear Dynamics: An Introduction for
Scientists and Engineers} (Oxford: Oxford University Press).

Krall, N. and Trivelpiece, A. (1986) \textit{Principles of Plasma Physics}
(San Francisco: San Francisco Press).

Koopman, B. and von Neumann, J. (1932), `Dynamical Systems of Continuous
Spectra,' \textit{Proceedings of the National Academy of Sciences}
\textbf{16}: 255-261.

Mackey, M. (1992) \textit{Time's Arrow: The Origins of Thermodynamic Behavior}
(Berlin: Springer-Verlag).

\_\_\_\_\_\_\_\_\_\_ (2002) `Microscopic Dynamics and the Second Law of
Thermodynamics', in C. Mugnai, A. Ranfagni and L. Schulman (eds.),
\textit{Time's Arrows, Quantum Measurement and Superluminal Behavior} (Rome:
Consiglio Nazionale Delle Ricerche), pp. 49-65.

Ord\'{o}\~{n}ez, A. (1998) `Rigged Hilbert Spaces Associated with
Misra-Prigogine-Courbage Theory of Irreversibility', \textit{Physica A}
\textbf{252}: 362-376.

Pathria, R. (1972) \textit{Statistical Mechanics} (Oxford: Pergamon Press).

Petrosky, T. and Prigogine, I. (1993) `Poincar\'{e} Resonances and the Limits
of Trajectory Dynamics', \textit{Proceedings of the National Academy of
Sciences USA} \textbf{90}: 9393-7.

\_\_\_\_\_\_\_\_\_\_\_\_\_\_\_\_\_ (1994) `Complex Spectral Representation and
Time-Symmetry Breaking', \textit{Chaos, Solitons \& Fractals} \textbf{4}: 311-359.

\_\_\_\_\_\_\_\_\_\_\_\_\_\_\_\_\_ (1996) `Poincar\'{e} Resonances and the
Extension of Classical Dynamics', \textit{Chaos, Solitons \& Fractals}
\textbf{7}: 441-497.

\_\_\_\_\_\_\_\_\_\_\_\_\_\_\_\_\_ (1997) `The Extension of Classical Dynamics
for Unstable Hamiltonian Systems', \textit{Computers \& Mathematics with
Applications} \textbf{34}: 1-44.

Petrosky, T., Prigogine, I. and Tasaki, S. (1991) `Quantum Theory of
Non-Integrable Systems', \textit{Physica A} \textbf{173}: 175-242.

Philippot, J. (1961) `Initial Conditions in the Theory of Irreversible
Processes', Physica 27: 490-6.

Prigogine, I. (1962) \textit{Non-Equilibrium Statistical Mechanics} (New York:
John Wiley \& Sons).

\_\_\_\_\_\_\_\_\_ (1980) \textit{From Being to Becoming: Time \& Complexity
in the Physical Sciences} (New York: W. H. Freeman).

\_\_\_\_\_\_\_\_\_ (1997) \textit{The End of Certainty: Time, Chaos, and the
New Laws of Nature} (New York: The Free Press).

Prigogine, I. and Balescu, R. (1959) `Irreversible Processes in Gases I.: The
Diagram Technique' \textit{Physica} \textbf{25}: 281-301.

Prigogine, I. and Petrosky, T. (1999) `Laws of Nature, Probability and Time
Symmetry Breaking' in Antoniou and Lumer (ed) \textit{Generalized Functions,
Operator Theory, and Dynamical Systems} (Boca Raton, FL: Chapman \& Hall/CRC Press).

Prigogine, I., Petrosky, T., Hasegawa, H And Tasaki, S. (1991) `Integrability
and Chaos in Classical and Quantum Mechanics', \textit{Chaos, Solitons \&
Fractals} \textbf{1}:3-24.

Primas, H. (1990) `Mathematical and Philosophical Questions in the Theory of
Open and Macroscopic Quantum Systems', in A. Miller (ed), \textit{Sixty-Two
Years of Uncertainty: Historical, Philosophical, and Physical Inquiries into
the Foundations of Quantum Mechanics} (New York: Plenum Press).

Roman, P. (1975) \textit{Some Modern Mathematica for Physicists and Other
Outsiders: An Introduction to Algebra, Topology, and Functional Analysis
Volume 2} (New York: Pergamon Press).

Tabor, M. (1989) \textit{Chaos and Integrability in Nonlinear Dynamics} (New
York: John Wiley \& Sons).

Treves, F. (1967) \textit{Topological Vector Spaces, Distributions and
Kernels} (New York: Academic Press).
\end{document}